# Investigating Corporate Social Responsibility Initiatives: Examining the case of corporate Covid-19 response

Meheli Basu[*a], Aniruddha Dutta[*b], Purvi Shah[c]

[a]Assistant Professor, Whitman School of Management, Syracuse University
(corresponding author: mbasu@syr.edu)

[b] Research Fellow, Smith School of Business, Queen's University

[a]Associate Professor, Foise School of Business, Worcester Polytechnic Institute

[*]*Denotes equal contribution*

June 30, 2023

**Abstract**

In today's age of freely available information, policy makers have to take into account a huge amount of information while making decisions affecting relevant stakeholders. While increase in the amount of information sources and documents increases credibility of decisions based on the corpus of available text, it is challenging for policymakers to make sense of this information. This paper demonstrates how policy makers can implement some of the most popular topic recognition methods, Latent Dirichlet Allocation and the Deep Distributed Representation method, and text summarization approaches, Word Based Sentence Ranking method and TextRank for sentence extraction method, to sum up the content of large volume of documents to understand the gist of the overload of information. We have applied popular NLP methods to corporate press releases during the early period (March to May 2020) and advanced period (June to August 2020) of Covid-19 pandemic which has resulted in a global unprecedented health and socio-economic crisis, when policymaking and regulations have become especially important to standardize corporate practices for employee and social welfare in the face of similar future unseen crises. The steps undertaken in this study can be replicated to yield insights from relevant documents in any other social decision-making context.

# 1. Introduction

Governments, policy makers, and regulators are increasingly involving stakeholders in solution-oriented policy making in a bid to be responsive to the needs of the society. They heavily rely on feedback on preferences and reviews from key stakeholders, such as companies, consumers, media, and non-governmental organizations (NGOs) in making regulations and policy frameworks [1-5]. These reviews, reports, and feedback constitute a huge collection of unstructured data, including the vast library of text available on the internet. One important domain of decision making for policy makers is Corporate Social Responsibility (CSR). CSR is defined broadly as 'a commitment to improve societal well-being through discretionary business practices and contributions of corporate resources' [6]. The growing importance of CSR can be attributed to a number of factors, including the pressure exerted on firms by policy makers, consumers, media, and NGOs to demonstrate high ethical standards [7]. CSR reporting was introduced originally in the form of environmental disclosures [8], and currently many governments worldwide require CSR disclosures regarding social aspects [9]. The required disclosure information is often in the form of economic, legal, ethical, and philanthropic responsibilities and drives of companies towards their stakeholders and the society at large [10]. Since the turn of the 21$^{st}$ century, CSR reporting has dramatically increased across industries [11-14]. According to the Governance and Accountability Institute, 2015 [15], the CSR reporting of S&P 500 companies has increased from about 20% in 2011 to over 70% in 2013, which helps inform stakeholders such as policymakers, consumers, investors.

In addition to CSR reports, the internet has numerous sources of information on welfare drives undertaken by companies. These include press releases, articles in newspapers and other publications, and discussions on social media platforms. So, the challenge is how do policy makers and other stakeholders make sense of the massive corpus of such unstructured textual data, be it corporate documents and press releases or independent reporting, for the innumerable firms spanning across several industries and sectors. In the CSR literature, researchers have long debated the effectiveness of such widely available textual data in communicating the gist of actual corporate engagement in CSR, and the challenges for policy makers and other stakeholders in discerning the information from the various sources of reporting on corporate societal contributions [16-19]. In this study we demonstrate how appropriate NLP techniques of text analysis can be applied to help policy makers and other stakeholders comprehend corporate social benefit engagement, in order to generate standard industry practices and monitor their implementation by corporations. In our study, we examine corporate engagement on social welfare practices during the Covid-19 pandemic, an unpredictable global circumstance when instituting rapid policies and appointing novel regulations by policy makers and governments become extremely critical.

According to the United Nations Development Program [20], beyond its impacts on health, the pandemic is also an unprecedented socio-economic crisis that has had deep economic impacts (undp.org article: COVID-19 pandemic Humanity needs leadership and solidarity to defeat the coronavirus). During Covid-19, companies have undertaken many societal drives and have regularly reported on their CSR efforts. For example, Tesla Inc. responded to the Covid-19 pandemic in order to benefit its stakeholders by purchasing and donating more than a thousand ventilators, offering its factory for manufacturing ventilators, and preparing its buildings and training its workforce to maintain social distancing. On the other hand, several employees tested positive for the coronavirus after the company disregarded shelter-in-place orders in order to restart manufacturing during the pandemic [21]. Companies have portrayed exemplary CSR behaviors and adapted their CSR policies and actions to current health, economic, and social needs [22]. On the other hand, they have also missed out on crucial elements that could have been managed better by learning from mistakes made by other organizations. The challenge is how to learn about what other organizations are doing to combat the impact of pandemic on their business, employees, consumers, communities in which they operate, and other key stakeholders. In order to inform stakeholders about their CSR initiatives and provide justifications for challenging decisions, companies have been regularly communicating with their respective stakeholders. Various companies have regularly generated content through press releases and blogs to update stakeholders on how they have been responding to the Covid-19 pandemic in a bid to alleviate organizational and social challenges. This has provided an interesting source of data that can help policy makers understand how corporations have done their part in fighting the pandemic.

This study exhibits the application of NLP techniques methods on corporate press releases of top 15 Nasdaq listed innovation firms about organizational, social, and employee welfare drives during the special case of the Covid-19 pandemic. We apply topic modeling methods of Latent Dirichlet Allocation and the Deep Distributed Representation method, and text summarization techniques of Word-Based Sentence Ranking method and TextRank for sentence extraction method to derive insights into the CSR practices of individual firms as well as the collective efforts of the top 15 innovation firms, in the wake of a pandemic. We demonstrate how the above techniques could be systematically leveraged by policy makers on similar unstructured textual data to frame policies for the benefit of society.

The rest of the paper is organized as follows. We first expand on the data collection process and discuss the NLP techniques that have been implemented on the text collected. This is followed by the results section, which details the analysis and insights from the implemented NLP methods.

This section also highlights how policymakers can use these methods for informed decision-making for social benefit. Next, we have a discussions and implementations section, discussing the overall findings, contributions and future research direction. We close with the acknowledgment section.

**2. Data & Methodology**

We systematically collected corporate press releases related to COVID-19 pandemic of top 15 NASDAQ listed companies, scraped from individual company websites during the time period March 2020 to August 2020. Nasdaq index is considered to investigate the CSR efforts of the most innovative technology US companies based on their market capitalization (Table 1) for systematic analysis; the technical methodology and framework is applicable to any such company classification (e.g., S&P 500, Fortune 500, Forbes Global 2000). At the time of data collection period, TECHNOLOGY sector comprised of 8 companies, CONSUMER SERVICES of 3 companies while the rest of the sectors constituted of one company each as shown in Table 1. The goal is to offer a standardized approach to policy makers and stakeholders, that can aid in the methodical analysis of text data, such as CSR reports, press releases, or other articles using popular NLP techniques to generate insights for framing transparent policies for societal benefit. The data collection time period represents the onset of COVID-19 pandemic period (March to May 2020) and later period as it progressed through 2020 (June to August 2020). The collected data is cleaned into tokens and phrases by parsing it through popular Python libraries NLTK and GENSIM and by using customized text mining algorithms. The text data is merged for two periods: (March – May 2020) period and (June - August 2020), to examine if there are differences in corporate social and employee welfare themes that corporates have focused on during the early and later pandemic periods. A distinct set of topics and associated keywords on social and employee benefit drives during the two periods of the pandemic are extracted from the data using Latent Dirichlet Allocation (LDA). Consequently, we employ the alternate model of Deep Distributed Representation method for topic modeling, which additionally also yields key sentences in the text documents that contain the keywords constituting the identified topics or themes in the collective text documents. This will provide insights for regulators beyond the main CSR themes during the pandemic, into identifying key document sentences that the algorithm aggregates to identify key CSR focus.

**Table 1: NASDAQ top 15 companies based on market capitalization.**

| Company | Sector |
|---|---|
| Apple Inc | Technology |
| Microsoft Corporation | Technology |
| Amazon.com Inc. | Consumer Services |
| Alphabet Inc. | Technology |
| Facebook Inc. | Technology |
| Tesla Inc. | Capital Goods |
| Nvidia Corporation | Technology |
| Paypal Holdings Inc. | Miscellaneous |
| Adobe Inc. | Technology |
| Comcast Corporation | Consumer Services |
| Netflix Inc | Consumer Services |
| Intel Corporation | Technology |
| Cisco Systems Inc. | Utilities |
| PepsiCo Inc | Consumer Non-Durables |
| Broadcom Inc. | Technology |

Finally, we perform text summarization on the press releases of an individual firm that usually provides an overload of information on past and ongoing CSR drives. Our results are intended to yield insights that the text summarization method can provide with, so policy makers can summarize large texts of social and employee benefit schemes that companies report over a given period of time. In our analysis, we summarize the large body of text on Covid-19 CSR press releases and blogs reported by Amazon.com during the period of March to August 2020. The insights retrieved from text summarization will facilitate policy makers to understand the summary of the bulk of self-reported information by private firms and to obtain a comprehensive understanding of firms' perspective on societal contributions. Our results also establish the fact that text summarization techniques could be systematically and effectively used for information retrieval from publicly available data sources. The following sections discuss the NLP methodologies employed in the present research.

**2.1.1 Latent Dirichlet Allocation Model**

Traditionally, researchers have used Latent Dirichlet Allocation (LDA) for topic modeling [23- 26] one of the most popular and interpretable generative probabilistic models for finding topics in text data. The LDA model describes documents as combinations of topics, each of which is described as a distribution of certain words. LDA generalizes Probabilistic Latent Semantic Analysis (PLSA) as [27], "Probabilistic latent semantic indexing", *Proceedings of the 22nd Annual International ACM* by adding a Dirichlet prior over document-topic and topic-word distributions. A low Dirichlet prior, $\alpha < 1$ for the topics in the documents mean that the documents are fairly separated in the topics present in them, while $\alpha > 1$ represents a more even mix of topics. A similar interpretation holds for the Dirichlet prior for words, $\beta$ representing how the topics are distributed amongst the words. We start with $\alpha$ and $\beta$ values as hyperparameters in LDA. Then the algorithm creates two Dirichlet distributions for the documents and topics and then generates documents based on those distributions. One of the several distributions created by the LDA algorithm generates the final document which closely estimates the words and associations found in the original document of interest.

LDA assumes that a document out of *D* documents is a probability distribution over *N* topics, where each topic is a probability distribution over all the different words in our vocabulary, *V*. Mathematically, the total probability of LDA model, i.e., the probability of a topic, *T* appearing in a given document and probability of a word, *W* appearing in the topics is given as follows:

$$P(W, T, \varpi, \phi; x, y) = \prod_{k=1}^{N} P(\phi_k; y) \prod_{d=1}^{D} P(\varpi_d; x) \prod_{m=1}^{M} P(T_{d,m}| \varpi_d) P(W_{d,t}|\phi_{d,m}) \quad (1)$$

where,

$\prod_{k=1}^{N} P(\phi_k; y)$ is the Dirichlet distribution (prior probability distribution) of *N* topics over words

$\prod_{d=1}^{D} P(\varpi_d; x)$ is the Dirichlet distribution of *D* documents over topics

$\prod_{m=1}^{M} P(T_{d,m}| \varpi_d) P(W_{d,t}|\phi_{d,m})$ is the product of two multinomial distributions: the probability that a topic appears in a given document and the probability that a word appears associated to a given topic;

$P(W, T, \varpi, \phi; x, y)$ is the resulting LDA model probability of generating a distribution of words close to that of the original text, by optimizing the values of *x* and *y* parameters.

One of the widely used metrics to interpret LDA model usefulness in generating topics is relevance, which measures the degree of popularity of a key word associated with a given topics, owing to it being specific to that topic as opposed to just being a highly frequent word in the document text. When we adjust relevance with a lower $\alpha$, we remove key words that is recognized simply by appearing frequently in the text by penalizing and removing such words. The weight adjusting term, $\alpha$ is optimized to identify more relevant keywords and thus determine meaningful topics in the given set of documents. Sievert and Shirley, 2014 [28] find that an $\alpha$ value in between 0.3 and 0.6 generates an optimal set of significant key words.

Mathematically, relevance is defined as follows:

$$R(W, T | \alpha) = \alpha \log(\chi_{TW}) + (1 - \alpha) \log\left(\frac{\chi_{TW}}{\delta_W}\right) \quad (2)$$

where,

$R$ is the relevance of key word, $W$ in association to topic, $T$

$W$ is a word in our vocabulary

$T$ is a topic produced by the LDA model

$\alpha$ is the adjusting parameter for the weight of a given word (for penalizing and removing words that appear very frequently throughout the text)

$\chi_{TW}$ is the probability of key word, $W$ appearing in topic, $T$

$\delta_W$ is the probability of key word, $W$ appearing in the full text (set of documents) being analyzed.

The discretization of topics is necessary to model the relationship between documents and words, especially for very large or unfamiliar datasets [29, 30]. Each topic produced by these methods is a distribution of word probabilities. Removing stop-words is an important pre-processing step since it is both language and corpus specific [31]. In the present study, frequent words in the corpus like Covid-19, Coronavirus are treated as stop-words since they were not very informative in the context of quantifying societal contributions.

**2.1.2 Deep Distributed Representation**

While the LDA model is appropriate to find topics, which are a latent representation of document-word distributions, the model fails to capture the sematic word relationships. To learn the distributed topic-word representations and to capture the context, we explore the *doc2vec* [32, 33) neural model scalable on large corpora which is otherwise not possible with LDA. In the era of artificial intelligence, these models are highly scalable and consistently perform better than non-neural methods [34-36]. The *doc2vec* model consists of a matrix $D_{i,j}$ where $i$ is the number of documents in the corpora and $j$ is the size of the desired vectors to be learned for each document. The document vector contains a word vector $\overrightarrow{w_k}$ of the order of dimension $j$ and is contained in each row of the matrix $D_{i,j}$. The model also uses a pre-trained matrix $W'_{i,j}$ which can be trained by existing models like *word2vec*. $W'_{i,j}$ is used as a context-word matrix; the vectors ($\overrightarrow{w_c}$) of which are used to predict the document's vector $j$ of $D_{i,j}$. The final prediction step is generated using the following operation:

$$softmax(\overrightarrow{w_k}.W'_{i,j}) \qquad (3)$$

where $\overrightarrow{w_k}$ is the context vector contained in $D_{i,j}$.

The neural network learns using back propagation and stochastic gradient descent to generate the highest probability $P(j|k)$ over the corpus of documents. This neural learning facilitates similar documents to cluster together in the semantic space. The density of similar documents indicative of an underlying topic can further be improved using existing spatial clustering techniques like Hierarchical Density-Based Spatial Clustering of Applications with Noise (HDBSCAN) [37-39] and dimensional reduction techniques like Uniform Manifold Approximation and Projection (UMAP) [40, 41]. In the present context of Covid-19 study, these methods are chosen for simplicity of implementation and interpretability of results. However, similar methods like T-distributed Stochastic Neighbor Embedding (t-SNE) [42] can also be used for similar purpose. We implement the *doc2vec* model along with HDBSCAN and UMAP to generate topics on the corporate corpus of press releases during Covid-19 pandemic to generate localized topics and associated keywords which are deep representation in the semantic space. Simultaneously we use topics and keywords to search documents which could be used by policy makers to locate the context of the topic and its associated keywords. This feature along with semantic representation makes this model more powerful and interpretable than other non-neural methods like LDA or PLSA [43].

**2.2 Text Summarization**

Text Summarization refers to condensation of a large body of text into a shorter version, while preserving its information content and overall meaning [44]. In today's fast-growing digital media age, text summarization has become an indispensable tool for summarizing and interpreting core themes from large number of text documents [45]. Extractive text summaries [46-48] are generated by extracting key text segments (sentences) from the text. Such extraction is based on statistical analysis of individual or mixed sentence features such as word frequency, location or cue words to locate the sentences to be extracted. For example, one of the earliest papers on text summarization by Luhn, 1958 [49] suggested to weight sentences of a document as a function of high frequency words, disregarding the very high frequency common words. In the summarization process, the "most important" content is condensed based on the "most frequent" or the "most favorably positioned" sentences. Features affecting the importance of sentences are selected, which are then assigned specific weights using weight learning method to determine final rank of each sentence. Top ranked sentences are then selected for the final text summary. There are many extractive summarization methods discussed in existing literature, such as term frequency-inverse document frequency (TF-IDF) method [50]; cluster-based method [51]; graph theoretic approach [52]; concept-obtained text summarization [53, 54]; neural networks approach to text summarization [55] automatic text summarization based on fuzzy logic [46, 56, 57] and query-based text summarization [58 - 60], among others. We have used two most common methods to summarize the text contained in the press releases of Amazon on CSR efforts during the Covid-19 period from March-August 2020. These are the sentence ranking method based on word frequency, and text rank for sentence extraction. Next, we discuss these two methods in details.

**2.2.1 Word Frequency based Sentence Ranking**

This method uses TF-IDF weights to rank sentences for text summarization. Key sentences are identified for summarizing, based on whether the sentences contain important words constituting themes of the clusters, based on the TF-IDF scores of the words. The process is defined as follows. Term Frequency-Inverse Document Frequency (TF-IDF) is a statistical metric that measures importance/weight of a word (term) in a corpus of text documents. Term Frequency is a ratio that measures the frequency of a given word appearing among all words in a document, as given by the equation below:

$$tf(w,d) = \frac{n_w}{\sum_i n_i} \qquad (4)$$

where, $tf(w,d)$ is the frequency of word, *w* appearing in document d; $n_w$ is the number of times word, *w* appears in the document; the sum of $n_i$ is the number of total words in the document. Inverse Document Frequency is the log of the ratio of number of all documents in the corpus, *D* to the number of documents, *$d_i$* with the word, *w*, defined by the following equation:

$$idf(w,D) = \log \frac{|D|}{|\{d_i \in D | w \in d_i\}|} \qquad (5)$$

Finally, tf-idf (w, d, D), which measures the frequency of a given word, *w* appearing among all words in a document, $d_i$ in the corpus of documents, *D*, is the product of term frequency and inverse document frequency, denoted by the equation below:

$$tf - idf(t, d, D) = tf(w, d) \cdot idf(w, D) \qquad (6)$$

After calculating the term frequencies, documents are represented using term frequency inverse document frequency (TF-IDF) scores of words [51]. This method of summarization uses the tf-idf score to identify key sentences containing important words to constitute the significant 'themes' within document clusters.

**2.2.2 TextRank for Sentence Extraction**

This method of sentence extraction from a document corpus ranks over text units, recursively computed based on information drawn from the entire text. This method is a derivative of the famous PageRank created by the Google co-founders. PageRank generates a matrix that calculates the probability that a user will move from one page to another. TextRank [61-63] generates a cosine similarity matrix for measuring the similarity of sentences to each other. A graph is then generated from this cosine similarity matrix and a derivative of the PageRank ranking algorithm to the graph to calculate scores for each sentence. In [61], the authors define the method in detail.

To apply TextRank, we first need to build a graph associated with the text, where the graph vertices represent sentences to be ranked. We define a "similarity" relation between two sentences, measured as a function of overlap of content of the two sentences. A sentence addressing certain concepts in a text, gives the reader a "recommendation" to refer to other sentences in the text addressing the same concepts, and thus a link can be drawn between any two similar sentence sharing similar content. Given two sentences $s_i$ and $s_j$, with sentence $s_i$ being represented by the set of $n_i$ words that appear in the sentence $s_i$ as: $w_1^i, w_2^i, ..., w_n^i$, the similarity between $s_i$ and $s_j$ is defined by the following equation:

$$similarity\ (si, sj) = \frac{|\{w_k\ |\ w_k \in s_i\ \&\ w_k \in s_j\}|}{\log(|s_i|) + \log(|s_j|)} \qquad (7)$$

The resulting graph is highly connected, with a weight associated with each edge measuring the strength of the connections established between sentence pairs in the text. After the ranking algorithm is run on the graph, sentences are sorted in reversed order of their score, and the top ranked sentences are selected for inclusion in the text summary.

In the next sections, we discuss data collection, analysis and results and follow up with a discussion on how policy makers can use the suggested content analysis methods to perform systematic exploration of corporate social drives and how they can impact the overall society that firms operate in.

**2.3 Results**

First, we perform LDA to analyze the top 5 topics and associated keywords appearing in the dataset of top NASDAQ firms, separately in the early pandemic period (March-May 2020) and in the later pandemic period (June-August 2020). Table 2 depicts the top 5 topics and associated keywords during the early pandemic period. We find that there are two topics, namely, "*healthcare efforts*" and "*relief efforts*", that highlight the social drives that the companies have undertaken during the early pandemic period. In healthcare efforts, some of the frequently appearing keywords are "*healthcare workers*", "*medical supplies*" and "*track cases*". This indicates that the top innovation companies have reported on their efforts to assist healthcare workers, their investment in assisting in medical supplies' acquirement and tracking covid cases. On the employee welfare front, companies reported to have prioritized employee safety and flexible working. Safety prioritization has been most often associated with keywords like "*enhanced clean*", "*social distancing*", "*temperature check*", "*essential items*" and "*personal protective equipment (PPE)*" within the early pandemic period of March-May 2020. Flexible working options appear to have included *extended leave* options, *hourly pay* and *work from home options*. Regarding internal operation during the early pandemic, firms have reported initiatives on accelerated research into rapid solutions and reported on leveraging resources and infrastructure to respond to operational interruptions due to Covid-19. So, one may find a balanced corporate focus on welfare of employees and company operations, as well as welfare schemes for the society during the early Covid-19 period.

**Table 2: Top 5 topics in Covid-19 related press releases of the top 15 Nasdaq firms during March-May 2020**

| Topic | Keywords | Coherence |
|---|---|---|
| Prioritize Safety | enhanced clean; mask; essential items; PPE; social distancing; temperature check | 0.452 |
| Flexible Working | hourly pay; extend leave; home; office; option | 0.405 |
| Accelerated Solution | innovation; respond; research; initiative; leverage resources; infrastructure | 0.434 |
| Healthcare Efforts | test; healthcare workers; medical supplies; collaborate; track cases; research; system deploy | 0.476 |
| Relief Efforts | fund; donation; charity; frontline workers; food; distribute; nonprofits; hunger; million; partner | 0.456 |

Table 3 contains the top 5 topics and associated keywords appearing on press release reports of the firms during the later pandemic period from June to August 2020. During this period, we see a lot of focus solely directed at the wellness of the society, as reflected in the top four topics appearing in company press releases: "*help community*", "*testing*", "*student learning*" and "*spreading information*". "*Help community*" appears to be associated with caring for employees and their families as well as communities and face global challenges during the pandemic period of June-August 2020, when the effects of the crisis were felt all over. Companies also have reported on helping in organizing access to free Covid-19 tests for patients and employees and appear to have helped with technology setup. "*Student learning*" also seems to have been a major corporate focus, as the pandemic progressed, and keywords such as students, education, remote access appear to be strongly correlated with such corporate initiatives. Corporations have also focused on their efforts to spread information during the pandemic in their June-Aug 20220 press releases. Keywords such as World Health Organization, mask, virus, platform, medical care and access to information have been frequently mentioned to highlight their efforts in establishing access to information on the virus and healthcare to communities. Finally, the reports also emphasize continued corporate focus on their internal operations, and online digital solutions along with customer service have been mentioned frequently in addition to employee safety and social distancing measures in their June-August reports. Thus, in the latter half of June-August 2020, we can take note that the corporate focus was hugely on social welfare initiatives.

**Table 3: Top 5 topics in Covid-19 related press releases of the top 15 Nasdaq firms during June-August 2020**

| Topic | Keywords | Coherence |
|---|---|---|
| Business Operation | online, open; employee safety; social distancing; customer service; digital; solutions | 0.380 |
| Help Community | crisis; families; employees; communities; global; challenges | 0.464 |
| Testing | free; patients; customers; access; testing; city; technology | 0.422 |
| Student Learning | learning; students; resources; education; communities; remote access | 0.511 |
| Spreading Information | world health organization; health; mask; global; virus; access to information; platform; medical care; monitor | 0.373 |

The above analysis helps demonstrate that policy makers can inform decision making on standardization of CSR practices by delving into systematic collection of press releases of companies, along with other valuable sources, and implement topic modeling to understand current best practices and their themes of industry leaders and other companies. To learn the semantic representation of topic and words, we train a neural network based on doc2vec model as discussed in section 2.1.2. A minimum threshold frequency of 10 is used to discard all words from the model during the training process. A window size which is the number of words left and right of the context word of 10 is used in the model. During the neural network training process, number of epochs implemented is optimized to a value of 10-50. For dimensional reduction process and minimum cluster size a value of 5 is set in the UMAP and HDBSCAN software. Table 4 depicts a subset of topics with keywords and corresponding cosine similarity score generated from the doc2vec model. Topics 3 and 4 generate keywords with relatively high cosine similarity scores as compared with topics 1 and 2.

Table 4: Topic keywords generated from doc2vec model along with cosine similarity score.

| Topics | Cosine Similarity Score |
|---|---|
| Topic1: Employees, safety, masks | 0.59, 0.48, 0.34 |
| Topic2: Devices, technology, connected | 0.47, 0.32, 0.29 |
| Topic3: Meals, children, community | 0.78, 0.74, 0.57 |
| Topic4: Education, virtual, digital | 0.75, 0.68, 0.62 |

Table 5: Combination keywords and document retrieval.

| Combination keywords | Score | Document |
|---|---|---|
| {masks, employees} (Document ID – 1622) | 0.703 | "Many health authorities now advise wearing non-medical masks and in some places masks are required for activities like taking pu--blic transportation or visiting a store and we have seen employees and businesses of all sizes working to fill this need." |
| {masks, employees} (Document ID – 1511) | 0.657 | "Frontline Supplies: Across the organization, we have donated and distributed masks, face shields, gowns, and other supplies to healthcare and frontline workers who need them most." |

Table 5 demonstrates how combination keywords could be used to determine the document ID to which it belongs along with retrieving the full document which in the present case represents a sentence. We retrieve the documents containing the keywords "*masks*" and "*employees*" from the corpus containing the press releases of top 15 Nasdaq companies related to Covid-19 along with the individual scores. The sentence corresponding to the document ID: 1622 contains both the keywords which fetches a higher score than the document ID: 1511. This deep continuous representation of topics and documents also demonstrate the fact that the topics which are being represented by the keywords are more localized than the non-neural techniques like LDA and PSLA. As a result, we illustrate the benefits of neural learning which comes at a computation cost. This technique can be used to retrieve information from a large corpus with specific keywords as inputs. Thus, the topic-keyword analysis in the deep semantic space provides a robust mechanism of big data information retrieval as an NLP technique which could be used for framing policies for societal benefit.

In the final step of our analysis of Covid-19 corporate welfare initiatives, we show how policy makers can make sense of the overall abstract of corporate initiatives from a broad range of publicly available documents, when organizations provide an overload of information. As an example of demonstration, we summarize the large body of text on Covid-19 CSR press release blogs reported by Amazon.com daily during the pandemic period from March 6th to August 30th, 2020. We summarize the text with the two summarization approaches discussed in section 2.2 of this paper. In the first approach of sentence scoring by word frequency, the text is summarized by scoring sentences in the documents based on the frequency of words making up the sentences. A dictionary of word frequency is generated for each word in the document as shown in Table 4 for the top 10 words. We assign weights to each word depending on the frequency of appearing in the documents. Using the weights assigned to each word, a score is created for each sentence (with normalization by dividing the absolute score by length of the sentence) as shown in Table 5 for the top 10 raw scores.

**Table 6: Top 10 frequent words in Amazon Covid-19 response reports from Mar-Aug'20**

| Word | Freq. | Word | Freq. |
|---|---|---|---|
| supported | 161 | food | 72 |
| donated | 124 | businesses | 66 |
| employees | 95 | relief | 49 |
| health | 92 | student | 44 |
| communities | 80 | nonprofit | 32 |

**Table 7: Top 10 scored sentences in Amazon Covid-19 response reports from Mar-Aug'20**

| Sentence | Raw Score |
|---|---|
| 'Amazon has created more than 4,000 full-time jobs across Massachusetts.' | 264 |
| 'Amazon's donation will provide more than 150,000 healthy, nutritious meals.' | 253 |
| ' Amazon collaborates with H.E.R. Tools like Amazon Chime will help interns connect with teammates, mentors, managers, and other interns.' | 248 |
| 'Meanwhile, Amazons corporate buildings are open and available to all employees.' | 215 |
| 'Amazon offers backup family care to 650,000 full- and part-time employees.' | 206 |
| 'Supporting communities in need throughout Massachusetts Amazon donated $100,000 to the Massachusetts COVID-19 Relief Fund.' | 201 |
| 'When you walk in, they take your temperature.' | 191 |
| 'Amazon Future Engineer will also welcome its class of nearly 100 interns, virtually.' | 189 |
| 'This comes after Amazons original donation of 8,200 laptops, valued at over $2 million.' | 183 |
| 'These tools will help small businesses use the cloud, communicate, and collaborate.' | 170 |

An average score of the sentences (*threshold*) is computed and multiplied with alpha ($\alpha$) value to determine the sentence cut-off score for qualifying sentences to be part of the final sentence summary. Thus, a sentence gets qualified to appear in the summary if the sentence score is greater than the value *threshold* * $\alpha$ . We vary the $\alpha$ values and find that $\alpha = 1.5, 2$ yield optimal interpretable summaries in terms of length. Following is the generated summary for $\alpha = 1.5$:

*"Amazon donated over 273,000 gallons of hand sanitizer to the Feeding America network of food banks across the U.S. Amazon's corporate buildings are open and available to all employees. Amazon donated 8,200 laptops, valued at over $2 million. Amazon offers backup family care to 650,000 full- and part-time employees. Respondents felt Amazon had adapted well to help keep people safe, according to a survey by Magid. Supporting communities in need throughout Massachusetts, Amazon donated $100,000 to the Massachusetts COVID-19 Relief Fund. Amazon's donation will help support dozens of Massachusetts-based nonprofits that serve communities disproportionately impacted by COVID-19. Amazon has created more than 4,000 full-time jobs across Massachusetts. Amazon's donation will provide more than 150,000 healthy, nutritious meals. For workers' safety, the fleet of robots perform more than 8,000 hours of cleaning each day. When you walk in, they take your temperature. This week, Amazon delivered personal protective equipment and supplies in support of Get Us PPE in Chicago. Amazon devices help connect elderly residents in care homes, hospital patients, and their families. Amazon will help small businesses use cloud, communicate, and collaborate. Amazon collaborates with H.E.R. Tools like Amazon Chime will help interns connect with teammates, mentors, managers, and other interns. Amazon Future Engineer will also welcome its class of nearly 100 interns, virtually. So far, 128 nonprofits have received funds to bolster frontline responses to the pandemic. Items donated include linens, towels, shelf-stable food, Amazon Devices, entertainment items, and other supplies. Additionally, we are subsidizing two full months of rent for tenants in Amazon-owned buildings."*

Following is the generated summary for $\alpha = 2$:

*"Amazon's corporate buildings are open and available to all employees. Amazon offers backup family care to 650,000 full- and part-time employees. Amazon Web Services supports COVID-19 High Performance Computing Consortium projects. Supporting communities in need*

*throughout Massachusetts Amazon donated $100,000 to the Massachusetts COVID-19 Relief Fund. Amazon has created more than 4,000 full-time jobs across Massachusetts. Amazon's donation will provide more than 150,000 healthy, nutritious meals."*

We observe that $\alpha = 2$ value for sentence score cut-off construction provides a shorter summary of Amazon.com reports from March-August 2020 on their Covid-19 response efforts, the $\alpha = 1.5$ value for cut-off score offers a more detailed summary of all Covid-19 measures aimed at societal and employee benefits, undertaken by Amazon during the 2020 pandemic period. Thus, the alpha value can be easily varied and optimized for the most optimal, comprehensive and interpretable summary of corporate measures, to summarize the gist of CSR policies by policy makers. If policy makers want to view larger and more detailed summaries of relevant documents for effective decision-making, they should choose a smaller $\alpha$ value so that more sentences can satisfy the smaller cut-off score value and can be included in the text summary, to create more complete summaries. On the other hand, if policy makers want to make decisions based on a smaller and more concise summary, they should choose a larger $\alpha$ value so that fewer sentences can satisfy the larger cut-off score value and the text summary consists of these fewer and more selective sentences from the large corpus of documents. Our analysis provides insights into how policy makers can summarize excess information from corporate documents reported by organizations in today's highly visible online platforms in the wake of mandated reporting transparency.

In the second text summarization approach, we use the TextRank algorithm, where cosine similarity between two given sentences is used as a transition probability and the similarity scores as shown in equation 7 are stored in a square matrix. The first step is to concatenate the text contained in the Amazon.com press releases and blogs reporting on the corporate pandemic response and then split the text into individual sentences. Next, we find vector representation (word embeddings) for every sentence, based on which the similarities between sentence vectors are calculated and stored in the matrix. The similarity matrix is then converted into a graph, with sentences as vertices and similarity scores as edges, for sentence rank calculation. Top sentences with highest cosine similarity scores to other sentences are selected. Finally, a specified number of top-ranked sentences form the final summary. We get the following summary of Amazon's Covid-19 relief efforts during March-August 2020, comprised of the top 5 sentences:

*In 2020, AWS helped the World Health Organization Academy (WHOA) launch an app to support health workers around the world in caring for patients infected by COVID-19. Amazon donated $5 million donation of devices globally to help healthcare workers, patients, students, and communities impacted by COVID-19.*

*This donation is part of Amazon's broader ongoing efforts to support customers, employees, and communities in the UK and around the world.*

*Delivering meals and donations to vulnerable populations in our HQ communities Amazon is using its delivery network to support the immediate needs of vulnerable populations in Seattle and the Washington D.C. area.*

*Amazon Web Services helped Crisis Cleanup, a software platform, to launch new websites to streamline volunteer and community relief responses to COVID-19.*

We get the following summary of Amazon's Covid-19 relief efforts during March-August 2020, comprised of the top 10 sentences:

*In 2020, AWS helped the World Health Organization Academy (WHO) launch an app to support health workers around the world in caring for patients infected by COVID-19.*

*Amazon Web Services helped launch Crisis Cleanup, a software platform, to streamline volunteer and community relief responses to COVID-19.*

*Amazon donated $5 million donation of devices globally to help healthcare workers, patients, students, and communities impacted by COVID-19.*

*Amazon's donations are part of the broader ongoing efforts to support customers, employees, and communities in the UK and around the world.*

*Delivering meals and donations to vulnerable populations in our HQ communities, Amazon is using its delivery network to support the immediate needs of vulnerable populations in Seattle and the Washington D.C. area.*

*Amazon Web Services (AWS) is supporting a new initiative from Volunteer Surge, a nonprofit consortium, to recruit, train, and deploy one million volunteer health workers.*

*The website is hosted on Amazon Web Services' cloud infrastructure, and we are proud to help ensure health care professionals around the world have access to this important resource during the COVID-19 pandemic.*

*Amazon is also partnering with Bellevue LifeSpring, an organization providing services to children and families, to help those in need obtain essential items during the outbreak.*

*The Spark Team, a nonprofit that provides software to support humanitarian events, is using AWS to help free medical clinics store patient data and manage volunteer sign-ups for COVID-19 test sites.*

*Amazon Web Services helps Propel, a company focused on building software to help low-income Americans improve their financial health, in developing the Fresh Electronic Benefit Transfer (EBT) app.*

We observe that the summaries generated by the two extractive approaches are different in their focus of Amazon's Covid-19 relief efforts. Both approaches yield informative summaries, and the level of informativeness of the summaries can be decided by policy makers based on which aspects are more important in decision-making. The best practice would be to explore both the algorithms although the TextRank algorithm is much more advanced and efficient [62].

**Discussion and Implications**

In this study, we have investigated two topic modeling approaches: LDA and a deep distributed representation algorithm, and two text summarization techniques: Word-Frequency based sentence ranking method and TextRank for sentence extraction method, to demonstrate how policymakers can sum up large corpus of text data implement informed and effect decision-making for standardizing practices for societal benefit, including corporate social responsibility practices. As context of text analysis, we examine the CSR activities of the NASDAQ top 15 companies during the Covid-19 pandemic from a diligently collected dataset of corporate press releases from March to August 2020 on how these companies responded to the challenges presented by the coronavirus pandemic. Applying NLP methods on corporate press releases and blogs on Covid-19 responses, enables understanding what companies have done in the wake of this crisis so that governments, policy makers, NGOs, media, consumers, and other key stakeholders can use this information to recommend what are the best practices and what more should be done for societal benefit and welfare.

In the Covid-19 CSR context, from LDA implementation on the press releases of the Nasdaq top 15 innovation companies, we find a collective balanced corporate focus on welfare of employees and company operations, as well as welfare schemes for the society during the early Covid-19 period of March-May 2020. In the latter half of the pandemic from June-August 2020, topics identified from the collective press releases signal that corporate focus was hugely on social welfare initiatives. We demonstrate that policy makers can inform decision making by delving into systematic collection of data and implement topic modeling to understand current best practices and their themes of industry leaders and other companies. Further we employ the alternate model of Deep Distributed Representation method for topic modeling, which additionally also yields key insights for regulators beyond the main CSR themes during the pandemic, into identifying key document sentences that the algorithm aggregates to identify key CSR focus.

We have also implemented two popular extractive text summarization approaches to demonstrate how policy makers can summarize the abstract of corporate initiatives from a broad range of publicly available documents, when organizations provide overload of information. We take the special case of Amazon, that regularly updated their Covid-19 relief efforts though blog and press release publications during the pandemic period from March 6th to August 30th, 2020. We summarize the text with the two summarization approaches of word frequency-based sentence ranking and TextRank sentence extraction methods. In the first approach of text summarization, we find that by simply altering the alpha value can alter the cut-off value of sentence selection score, and can yield desired length of summary. If policy makers want to view larger and more detailed summaries of relevant documents for effective decision-making, they should choose a smaller optimal $\alpha$ value. Whereas, if policy makers want to make decisions based on a smaller and more concise summary, they should simply choose a larger optimal $\alpha$ value. The second text summarization approach, TextRank algorithm extracts summary sentences from the large volume of text documents based on the cosine similarity of each sentence to others in the text. We observe that the summaries generated by the two extractive approaches are different in their focus of Amazon's Covid-19 relief efforts. The level of usefulness of the different informative summaries generated by the two extractive approaches, can be decided by policy makers based on which decision-making aspects are better reflected in the summaries.

These findings have implications for various stakeholders. Policy makers can use the data analysis approaches presented in this paper for text analysis of CSR reports, press releases, newspaper publications, blogs and other sources of relevant text documents. The information they generate using these methods will help them understand what individual companies are doing to respond to the challenges hurled by the pandemic. Policy makers can further apply these methods on aggregated data at the level of the industry and the sector to standardize industry specific best practice expectations. This knowledge will further aid the policy making process in making recommendations about (1) what are the best practices by industry leaders, (2) what practices companies are neglecting, (3) what they should be doing, and (4) what they should not be doing in the event of a pandemic or any other crisis situation. Furthermore, this data collection and analysis method is also helpful to the media, consumers, and NGOs. The information gained from the NLP text analysis educates consumers and helps them decide whether to do business with the company or not based on their CSR behaviors during the pandemic. It also provides information that media and NGOs could use to influence policy making, empower consumers, and encourage companies to do the "right things" for social and employee benefits. Finally, these findings have implications for the companies as well. Companies are sailing through uncharted waters during the Covid-19 pandemic, that involves a steep learning curve for organizational decision-making. It will benefit companies by saving response time and resources if they can learn from each other's strategies, experiences and mistakes. They can share their stories, educate their consumers and employees, persuade media and governments that they are doing their best to mitigate the adverse impact of the pandemic, strategically collaborate and partner with other organizations as well as governments and NGOs for the greater good. Moreover, through such text analysis techniques, companies can learn from other organizations' CSR initiatives and propose, and adopt industry and sector level best practices to survive and succeed in a pandemic and at the same time ensure societal and environmental well-being.

Though we use only one data source of corporate press releases about CSR engagement during the pandemic as a case, the method we demonstrate in this study is also generalizable to other types of data sources and this analysis could be used (1) for various types of organizations such as corporate or NGOs and (2) in various contexts, beyond crisis situations like a pandemic. For future research, gathering data from multiple sources such as media reports, news articles, and consumer generated data on social media can provided a more holistic and unbiased view of corporate social practices. The same NLP techniques can be applied to the analyze these forms of data. In addition, sentiment analysis techniques could be used for social media data and opinion mining and belief and discourse analysis could be used for media reports and news articles. These methods could also be applied in the future to different contexts such as (1) Company perspective: Studying the impact of CSR initiatives on a company's reputation, in generating positive word-of-mouth, in managing consumer reactions to brand crisis situations, and in developing brand loyalty and commitment amongst consumers and employees, (2) Public policy perspective: Investigating the CSR engagement of companies on a regular basis, standardizing public policy on challenging industry-specific aspects of CSR, using CSR information to empower consumers and employees to make informed choices, and (3) Stakeholder perspective: Enabling key stakeholders make important decisions such as (a) Should I buy products from this organization?; (b) Should I invest in this organization?; (c) Should I work for this organization?; (d) Should I do business with this organization?; and so on. Overall, the NLP text analysis methods we demonstrated in this study will serve as a useful tool for policy makers and key stakeholders in understanding how to standardize best practices and institute regulations for the overall welfare of societies. Regulators can make informed decisions based on the insights drawn from the text analysis, and emerge successful with win-win solutions for all stakeholders impacted.

**Acknowledgements**

The authors thank the Library and Information Technology Services at Syracuse University for providing computing cluster for the extensive processing and analysis of the large volume of documents collected for detailed content analysis implementation. This research did not receive any specific grant from funding agencies in the public, commercial, or not-for-profit sectors.